\documentclass[preprint,amsmath,amssymb,superscriptaddress]{revtex4}
\usepackage{graphicx,color}
\usepackage{amssymb}
\usepackage{amsmath}
\usepackage{latexsym}
\usepackage{amsxtra}
\usepackage{amscd}
\usepackage{dcolumn}
\usepackage{bm}

\graphicspath{{./}}
\let\oldsqrt\sqrt
\def\sqrt{\mathpalette\DHLhksqrt}
\def\DHLhksqrt#1#2{\setbox0=\hbox{$#1\oldsqrt{#2\,}$}\dimen0=\ht0
\advance\dimen0-0.2\ht0
\setbox2=\hbox{\vrule height\ht0 depth -\dimen0}%
{\box0\lower0.4pt\box2}}

\usepackage{hyperref}

\begin{document}

\title{Identification of the slow E3 transition $^{136m}$Cs $\rightarrow$ $^{136}$Cs with conversion electrons}

\author{K.~Wimmer}\altaffiliation[present address ]{National Superconducting Cyclotron Laboratory, Michigan State University, East Lansing, Michigan 48824, USA, wimmer@nscl.msu.edu}\affiliation{Fakult\"at f.~Physik, Ludwig-Maximilians-Universit\"at M\"unchen, 85748 Garching, Germany}
\affiliation{Physik Department E12, Technische Universit\"at M\"unchen, 85748 Garching, Germany}
\author{U.~K\"oster}\affiliation{Institut Laue Langevin, 38042 Grenoble Cedex 9, France}
\author{P.~Hoff}\affiliation{Department of Chemistry, University of Oslo, 0315 Oslo, Norway }
\author{Th.~Kr\"oll}\affiliation{Physik Department E12, Technische Universit\"at M\"unchen, 85748 Garching, Germany}
\affiliation{Institut f\"ur Kernphysik, Technische Universit\"at Darmstadt, 64289 Darmstadt, Germany}
\author{R.~Kr\"ucken}\affiliation{Physik Department E12, Technische Universit\"at M\"unchen, 85748 Garching, Germany}
\author{R.~Lutter}\affiliation{Fakult\"at f.~Physik, Ludwig-Maximilians-Universit\"at M\"unchen, 85748 Garching, Germany}
\author{H.~Mach}\affiliation{Department of Nuclear and Particle Physics, Uppsala University, 75121 Uppsala, Sweden }
\author{Th.~Morgan}\affiliation{Fakult\"at f.~Physik, Ludwig-Maximilians-Universit\"at M\"unchen, 85748 Garching, Germany}
\author{S.~Sarkar}\affiliation{Department of Physics, Bengal Engineering and Science University, Shibpur, Howrah 711103, India}
\author{M.~Saha~Sarkar}\affiliation{Nuclear and Atomic Physics Division, Saha Institute of Nuclear Physics, Kolkata 700064, India}
\author{W.~Schwerdtfeger}\affiliation{Fakult\"at f.~Physik, Ludwig-Maximilians-Universit\"at M\"unchen, 85748 Garching, Germany}
\author{P.~C.~Srivastava}\altaffiliation[present address ]{Physical Research Laboratory, Ahmedabad 380 009, India}\affiliation{Grand Acc\'el\'erateur National d'Ions Lourds, CEA/DSM-CNRS/IN2P3, BP 55027, 14076 Caen Cedex 5, France}
\affiliation{Nuclear Physics Group, Department of Physics, University of Allahabad 211002, India}

\author{P.~G.~Thirolf}\affiliation{Fakult\"at f.~Physik, Ludwig-Maximilians-Universit\"at M\"unchen, 85748 Garching, Germany}
\author{P.~Van~Isacker}\affiliation{Grand Acc\'el\'erateur National d'Ions Lourds, CEA/DSM-CNRS/IN2P3, BP 55027, 14076 Caen Cedex 5, France}

\begin{abstract}
We performed at ISOLDE the spectroscopy of the decay of the $8^-$ isomer in $^{136}$Cs by $\gamma$ and conversion-electron detection. For the first time the excitation energy of the isomer and the multipolarity of its decay have been measured. The half-life of the isomeric state was remeasured to $T_{1/2}=17.5(2)$~s. This isomer decays via a very slow 518~keV E3 transition to the ground state. In addition to this, a much weaker decay branch via a 413~keV M4 and a subsequent 105~keV E2 transition has been found. Thus we have found a new level at 105~keV with spin $4^+$ between the isomeric and the ground state.
The results are discussed in comparison to shell model calculations.
\end{abstract}



\maketitle

\section{Introduction}\label{sec:intro}
Although being close to stability, at present only two levels are known in $^{136}$Cs ($T_{1/2}=13.16$~d), namely the spin $5^+$ ground state and an isomer with a half-life of $T_{1/2}=19(2)$~s~\cite{ravn} and spin $8^-$~\cite{thibault}. 
The excitation energy of the $8^-$ state and its decay properties were not reported in literature prior to our measurement. 
Further interest in the structure of this nucleus comes from the fact that the $^{136}$Xe $\rightarrow$ $^{136}$Ba decay is a candidate for the observation of neutrinoless double $\beta$ decay~\cite{akimov05}, as the $\beta$ decay of $^{136}$Xe to $^{136}$Cs is energetically forbidden. 
For the calculation of $0\nu\beta\beta$ half-lives wave functions of all excited states of the intermediate nucleus are needed~\cite{suhonen98}. To test such calculation for the $^{136}$Xe neutrinoless double $\beta$ decay more information on the level scheme of $^{136}$Cs is required.

In this paper we report on the first spectroscopy of the decay of the $8^-$ isomeric state in $^{136}$Cs. The lifetime of the isomer as well as the multipolarities of the decays to the ground state and an excited state have been measured. 
The results are compared with shell model calculations.

\section{Experiment and Results}

Conversion-electron spectroscopy is an excellent tool to determine transition multipolarities, since the conversion coefficient $\alpha$ strongly depends on the multipolarity.
In order to determine the energy and the multipolarity of the transition, $\gamma$ rays and conversion electrons from the isomeric level have been measured at the ISOLDE facility at CERN. $^{136m}$Cs was produced by 1.4~GeV proton beam induced fission in a uranium carbide target. The reaction products diffused from the 2000~$^\circ$C hot target to a 2000~$^\circ$C hot tungsten ionizer, where cesium was surface ionized. After extraction and acceleration to 40~keV the ISOLDE high resolution separator was used to select an $A=136$ beam. Due to the high production rate of $^{136}$Cs, the primary beam intensity was reduced to one proton pulse every 16.8~s with a rather low intensity of $3\cdot10^{12}$~p/pulse. The Cs beam was sent to an aluminum catcher foil. In order to determine the multipolarity of the transition, the conversion coefficients for the K, L and M shell have been measured. The $\gamma$ decay was measured with a HPGe detector (photo peak efficiency $\epsilon_{\rm ph}(518\rm ~keV)=0.187~\%$ for the given geometry). Simultaneously, conversion electrons were registered in a liquid N$_2$ cooled Si(Li) detector (distance to target foil 91~mm, efficiency 0.43~\%). 
The efficiencies of both detectors have been determined using standard $\gamma$ ray and electron sources of $^{207}$Bi and $^{152}$Eu.
From the relative intensities of $\gamma$ and conversion-electron radiation, the conversion coefficients $\alpha_{\rm K,L,M,...}=I_{\rm e}^{\rm K,L,M,...} / I_\gamma$ have been determined. 

\subsection{The 518~keV transition}\label{sec:518}

\begin{figure}[h]
\includegraphics[width=0.48\textwidth]{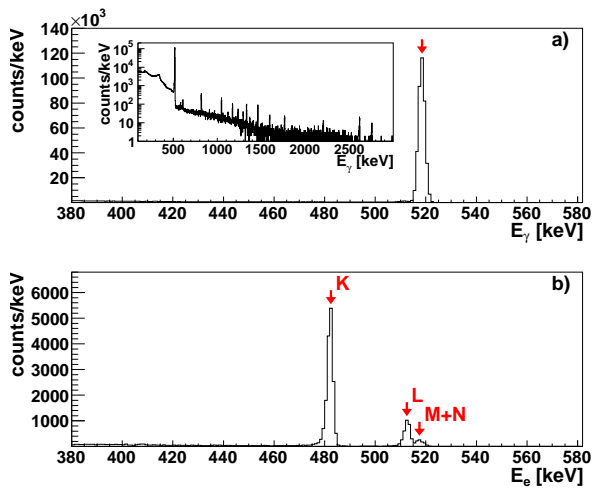}
\caption{a) $^{136m}$Cs $\gamma$-ray spectrum measured in 25~min. The 518~keV $8^-\rightarrow 5^+$ transition leading to the ground state of $^{136}$Cs is the most intense line observed. The inset shows the total $\gamma$-ray spectrum. The small lines originate from the $\beta$ decay of $^{136}$Cs to $^{136}$Ba and from natural radioactivity.
b) Corresponding conversion-electron energy spectrum. Conversion-electron lines from the K-shell (binding energy 36~keV), the L-shell (6~keV) and the M-shell (1~keV) can be distinguished.}
\label{518_fig}
\end{figure}

Fig.~\ref{518_fig} shows the $\gamma$-ray and conversion-electron spectra. 
From the measured $\gamma$-ray energy we deduce the previously unknown excitation energy of the $8^-$ state to be $517.9(1)$~keV. 
Table~\ref{518_alpha} shows the results of the measurement together with theoretical Band-Raman internal conversion coefficients~\cite{bricc} for the two possible multipolarities E3 and M4.
\begin{table*}[h]
\begin{tabular}{|r|r|l|ll|}
\hline
 shell & E$_{\rm e}$ [keV] & $\alpha_{\rm exp}$ & \multicolumn{2}{l|}{$\alpha_{\rm theo}$}\\
 & & & E3 & M4\\
\hline
K & 482 & $1.84(4)\cdot10^{-2}$ & $1.89(3)\cdot10^{-2}$ & $2.19(3)\cdot10^{-1}$ \\
L & 512 & $3.73(10)\cdot10^{-3}$ & $3.74(5)\cdot10^{-3}$ & $4.30(6)\cdot10^{-2}$ \\
M & 517 &  & $7.89(11)\cdot10^{-4}$ & $9.22(13)\cdot10^{-3}$ \\
N & 518 &  & $1.64(2)\cdot10^{-4}$ & $1.94(3)\cdot10^{-3}$ \\
M + N & 517 & $9.35(38)\cdot10^{-4}$ & $9.53(11)\cdot10^{-4}$ & $1.12(1)\cdot10^{-2}$ \\
\hline
\end{tabular}
\caption{Theoretical~\cite{bricc} and experimental conversion coefficients for the 518~keV transition in $^{136}$Cs. Since the binding energies for electrons in the M (1~keV) and N (0.2~keV) shell cannot be resolved in the experiment, only the sum of $\alpha_{\rm M}$ and $\alpha_{\rm N}$ can be measured. The conversion coefficients for M4 transitions are about an order of magnitude larger than for E3 transitions, thus the unknown multipolarity of the $8^-\rightarrow 5^+$ transition can be determined easily.}
\label{518_alpha}
\end{table*}
The measured conversion coefficients agree within the error with the calculated values for an E3 transition, thus we conclude that the 518~keV $8^-\rightarrow 5^+$ transition is of pure E3 character.

In a separate experiment the half-life of the isomeric state was remeasured with a HPGe detector placed at the ISOLDE monitoring tape station. The half-life of the 518~keV $8^-$ state amounts to $T_{1/2}=17.5(2)$~s. This is in agreement with the two previous measurements of $19(2)$~s~\cite{ravn} and $17(2)$~s~\cite{baba} but a factor of 10 more precise.

From the now known transition energy, multipolarity and half-life, 
the transition strength can be calculated. The long half-life $T_{1/2}=17.5(2)$~s results in a very small $B({\rm E3};\, 8^-\rightarrow 5^+)=6.97(8)\cdot 10^{-3}$~$e^2$fm$^6$ value. With only $6.35(7)\cdot 10^{-6}$~W.u., the $^{136m}$Cs $\rightarrow$ $^{136}$Cs transition is by far the slowest E3 transition known in this mass range. In the entire chart of nuclides there are only two known E3 transitions that are still slower, namely the decays of the $K$-isomers $^{177m}$Lu and $^{179m2}$Hf. There may exist other slow E3 transitions which are not observable due to competing faster transitions.

\subsection{The 413 keV transition}\label{sec:413}

\begin{figure}[h]
\centering
\includegraphics[width=0.48\textwidth]{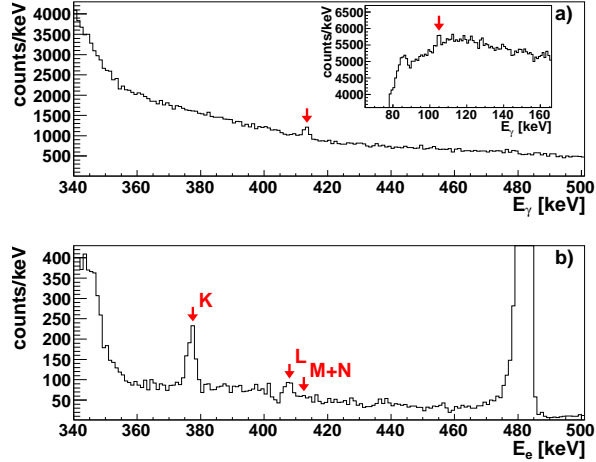}
\caption{a) $^{136}$Cs $\gamma$-ray spectrum around 400~keV. The transition at 413~keV is about a factor of 800 weaker than the 518~keV ground-state transition. The inset shows the 105~keV transition. 
b) Corresponding conversion-electron energy spectrum. }
\label{413_fig}
\end{figure}

In addition to the 518~keV ground-state transition, a 413.1(3)~keV $\gamma$ transition which is a factor of 800 weaker, and the corresponding conversion electrons have been found (Fig.~\ref{413_fig}). 
By measuring the conversion coefficients of this line, we conclude that this transition with $\alpha_{\rm K}=0.49(5)$ is an M4 transition (see Table~\ref{413_alpha}).
\begin{table*}[h]
\begin{tabular}{|r|r|l|ll|}
\hline
 shell & E$_{\rm e}$ [keV] & $\alpha_{\rm exp}$ & \multicolumn{2}{l|}{$\alpha_{\rm theo}$}\\
 & & & M4 & E5\\
\hline
K       & 377 & $0.490(47)$             & $0.543(80)$               & $0.306(5)$             \\
L       & 407 & $0.122(15)$             & $0.122(20)$               & $0.183(3)$             \\
M       & 412 &                             & $2.65(4)\cdot10^{-2}$   & $4.13(6)\cdot10^{-2}$ \\
N       & 413 &                             & $0.56(1)\cdot10^{-2}$   & $0.84(1)\cdot10^{-2}$ \\
M + N   & 412 & $3.1(11)\cdot10^{-2}$   & $3.21(4)\cdot10^{-2}$   & $4.97(6)\cdot10^{-2}$ \\
\hline
K/L     &     & $4.0(6)$               & $4.46(9)$               & $1.67(4)$               \\
L/(M+N) &     & $4.0(15)$               & $3.80(10)$              & $3.68(9)$               \\
\hline
\end{tabular}
\caption{Theoretical~\cite{bricc} and experimental conversion coefficients for the 413~keV transition in $^{136}$Cs. In addition to the conversion coefficients also the intensity ratios of conversion electrons K/L and L/(M+N) are given. This allows to distinguish between an M4 (K/L=4.46) and an E5 (K/L=1.67) transition. The K conversion coefficients for other multipolarities are at least a factor of 5 smaller (E4 transition) or a factor of 3 larger (for M5 transitions).}
\label{413_alpha}
\end{table*}
We also observed 105~keV $\gamma$ rays and coincidences between these and 377~keV K conversion electrons. This suggests that we have found a new level between the isomeric state at 518~keV and the ground state. Conversion electrons from the 105~keV transition have not been observed due to the detection threshold.
Under the assumption that no other levels exist, such that only the 413~keV transition can occur before or after a 105~keV transition, one can also determine the multipolarity of the 105~keV transition. Under the assumption of a cascade of 413 and 105 keV transitions, the yields shown in Table~\ref{mult_105} are consistent with an E2 transition only.
\begin{table*}[h]
\begin{tabular}{|r|r|r|r|r|r|}
\hline
 E$_{\gamma}$ [keV] & counts & $\epsilon_\text{ph}(\text{E}_{\gamma})$ & $\alpha_\text{tot}$~\cite{bricc}& mult. & $\text{counts} / \epsilon_\text{ph} \cdot (1+ \alpha_\text{tot})$ \\
\hline
413 & 567(48) & 0.00226 & 0.697(10) theo & M4 & $4.24(37)\cdot10^5$ \\
    &         &         & 0.642(119) exp & M4 & $4.13(46)\cdot10^5$ \\
\hline
105 & 830(58) & 0.00535 & 0.191(3)  & E1 & $1.85(13)\cdot10^5$ \\
    &         &         & 0.782(11) & M1 & $2.77(20)\cdot10^5$ \\
    &         &         & 1.550(22) & E2 & $\mathbf{3.96(29)\cdot10^5}$ \\
    &         &         & 7.47(11)  & M2 & $1.32(9)\cdot10^6$ \\
    &         &         & 17.32(25) & E3 & $2.84(21)\cdot10^6$ \\
    &         &         & 60.7(9)   & M3 & $9.58(71)\cdot10^6$ \\
\hline
\end{tabular}
\caption{Transition rates for the 413~keV and the 105~keV transition assuming different multipolarities for the 105~keV transition. The yield from the 413~keV line can only be reproduced using the conversion coefficient for E2 transitions for the 105~keV transition.}
\label{mult_105}
\end{table*}

\section{Proposed level scheme}\label{sec:levelscheme}

\begin{figure}[h]
\includegraphics[width=0.48\textwidth]{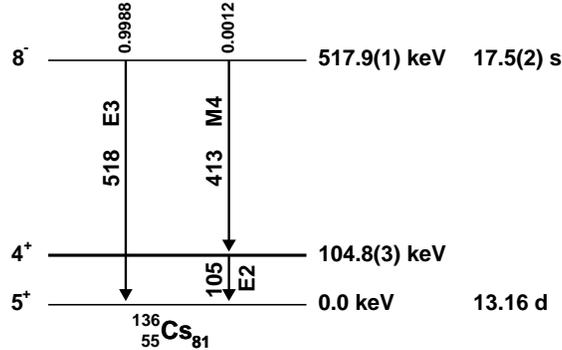}
\caption{Proposed level scheme of $^{136}$Cs. Previously known were only the spins of the ground state and the isomeric state as well as the half-life of $^{136}$Cs.}
\label{level}
\end{figure}

Since the 518~keV transition is the by far strongest line in the spectrum and no higher-lying lines of similar strength have been found, we conclude that this is the $8^-\rightarrow 5^+$ ground-state transition. Thus the $17.5(2)$~s $8^-$ isomeric state in $^{136}$Cs is located at 517.9(1)~keV. It decays via an E3 transition to the ground state. In addition to this $8^-\rightarrow 5^+$ transition an M4 transition with an energy of 413.1(3)~keV exists. This transition is a factor of about 800 weaker. The yield in the 105~keV transition line can only consistently be described by an E2 transition. This E2 transition is very unlikely to be in competition with the very slow E3 transition from the $8^-$ state, since it is much weaker. Thus we place the 413~keV transition above the 105~keV transition, giving rise to a previously unknown level at 104.8(3)~keV. The excitation energy of the level has been determined from the energy difference of the two transitions depopulating the 518~keV level.
The isomeric state decays with 99.9~\% directly and with 0.12~\% via a cascade of an M4 transition to the level at 105~keV followed by an E2 transition to the ground state. The intensities and multipolarities of the three transitions indicate a spin of the 105~keV level of $4^+$ or $5^+$. 
Assuming $J^\pi = 5^+$ for the 105~keV level, an M4 transition for the 413~keV transition with $\Delta I=3$ would be strongly suppressed compared to an alternative E3 transition, which is excluded by the conversion coefficient measurement. 
Therefore, we assign spin and parity $J^\pi = 4^+$ to the level at 105~keV.
Fig. \ref{level} shows the proposed level scheme of $^{136}$Cs.

\section{Shell-model interpretation}
\label{s_calc}
The standard shell-model description of $^{136}$Cs considers
the five protons and 31 neutrons
in the 50--82 valence shell
composed of the orbits
$1g_{7/2}$, $2d_{5/2}$, $1h_{11/2}$, $3s_{1/2}$, and $2d_{3/2}$.
Since 31 neutron particles in this shell
are equivalent to one neutron hole,
the problem is manageable with existing shell-model codes.
Before doing the full calculation,
it is instructive to obtain single-particle estimates
of the matrix elements
that enter into the decay of the $8^-$ isomer.
This is illustrated here for the E3 transition
from the $8^-$ isomer to the $5^+$ ground state. 

A reasonable simplification of the full calculation is to assume that,
both for the $5^+$ ground state and the $8^-$ isomer,
the five protons are in a seniority $v=1$ state
in the $\pi g_{7/2}$ and/or $\pi d_{5/2}$ orbits, 
that is, four of them couple to $J=0$ pairs and one remains unpaired.
The two proton pairs are then treated as inactive spectators
and the electromagnetic transition effectively takes place
between one-proton-particle--one-neutron-hole states.
The components of the wave functions
determining the electromagnetic decay from $8^-$ to $5^+$
involve a proton in the $\pi g_{7/2}$ orbit in both initial and final state,
while the configuration of the neutron-hole is $(\nu h_{11/2})^{-1}$ for the isomer
and $(\nu d_{5/2})^{-1}$ or $(\nu g_{7/2})^{-1}$ for the ground state.

It is a matter of standard angular momentum coupling~\cite{Talmi93}
to derive the result
\begin{eqnarray}
&&\langle\pi g_{7/2}(\nu d_{5/2})^{-1};5^+||
\hat T({\rm E}3)
||\pi g_{7/2}(\nu h_{11/2})^{-1};8^-\rangle
\nonumber\\
&&\qquad=
\frac{45}{4}\sqrt{\frac{85}{11\pi}}\,b^3e_\nu
\approx17.64\,b^3e_\nu,
\label{e_e3mat4}
\end{eqnarray}
where $b$ is the oscillator length.
With the first-order estimates $b\approx1.00A^{1/6}$~fm~\cite{Brussaard77}
and $e_\nu\approx0.5e$,
one finds a $B({\rm E3};8^-\rightarrow5^+)$ value of 622~$e^2$fm$^6$.
Not surprisingly, the single-particle estimate of the E3 transition is 0.57~W.u.,
hence of the order of one Weisskopf unit.
This leads to a fast decay 
($T_{1/2}=0.2$~ms),
which disagrees
with the observed half-life of $17.5(2)$~s of the $8^-$ isomer.
For the other component of interest in the $5^+$ ground state,
involving $(\nu g_{7/2})^{-1}$,
one finds a somewhat smaller estimate
\begin{eqnarray}
&&\langle\pi g_{7/2}(\nu g_{7/2})^{-1};5^+||
\hat T({\rm E}3)
||\pi g_{7/2}(\nu h_{11/2})^{-1};8^-\rangle
\nonumber\\
&&\qquad=
15\sqrt{\frac{51}{286\pi}}\,b^3e_\nu
\approx3.57\,b^3e_\nu,
\label{e_e3mat5}
\end{eqnarray}
but still too large to account for the observed long half-life of the $8^-$ isomer.
In order to try to explain this discrepancy,
a full shell-model calculation must be undertaken.

For the full shell-model calculation
we have used the interaction SN100PN
taken from Brown {\it et al.}~\cite{Brown05},
who applied it to the description of magnetic moments
of nuclei in the $^{132}$Sn region.
This interaction has four parts,
corresponding to the neutron--neutron, neutron--proton, and proton--proton
pieces of the nuclear force
and the Coulomb repulsion between the protons.
The single-particle energies for the neutrons are
$-10.610$, $-10.290$, $-8.717$, $-8.716$, and $-8.816$~MeV
for the $g_{7/2}$, $d_{5/2}$, $d_{3/2}$, $s_{1/2}$, and $h_{11/2}$, respectively,
and those for the protons are 0.807, 1.562, 3.316, 3.224, and 3.603~MeV.
With these energies,
the excitation energies of the single-particle states
in $^{131}$Sn and $^{133}$Sb,
the neutron separation energy of $^{132}$Sn,
and the proton separation energy of $^{133}$Sb
are reproduced exactly.
There is some uncertainty concerning
the location of the $s_{1/2}$ single-particle state in $^{133}$Sb;
if the corresponding energy is varied around the value adopted here (3.224~MeV),
the results shown below do not change significantly.
The shell-model Hamiltonian
is diagonalized with the code {\tt NuShell}~\cite{Brownun}.

\begin{figure}
\begin{center}
\includegraphics[width=0.48\textwidth]{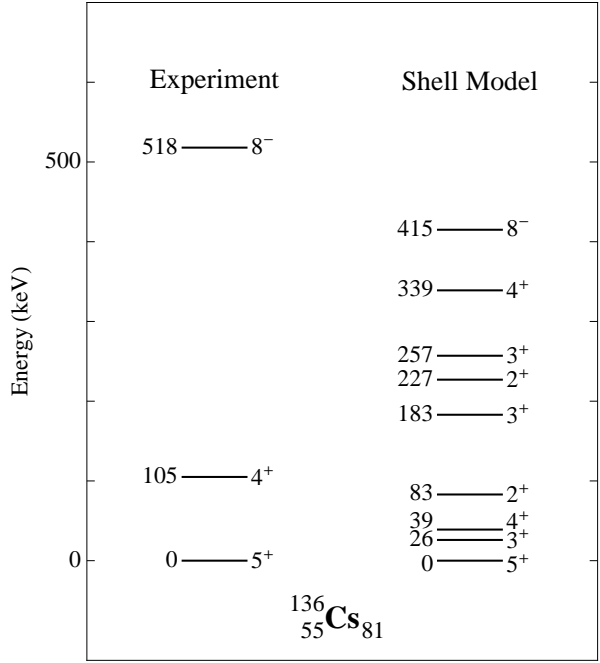}
\caption{
The energy levels observed in $^{136}$Cs
compared with the results of a shell-model calculation
with the SN100PN interaction.
Excitation energies (in keV) are indicated to the left of the levels.}
\label{f_cs136e}
\end{center}
\end{figure}

The results of the calculation are shown in Fig.~\ref{f_cs136e}.
The shell model correctly predicts a $J^\pi=5^+$ ground state 
and a $J^\pi=8^-$ state at 415~keV,
not too far from the experimental value of 518~keV.
Seven levels are predicted between the $5^+$ and $8^-$ states,
two of them with $J^\pi=4^+$
are possible candidates for the additional level
observed between the isomer and the ground state.

The analysis of the wave functions obtained
for the $5^+$ ground state and the $8^-$ isomer
is of interest to understand the E3 decay.
The dominant component (49.9\%) of the ground state
has the structure $|(\pi g_{7/2})^5(\nu d_{3/2})^{-1};5^+\rangle$
with five protons in $\pi g_{7/2}$
and a neutron-hole in $\nu d_{3/2}$.
The $8^-$ is exclusively composed
of a neutron-hole in $\nu h_{11/2}$
with as most important components
$|(\pi g_{7/2})^5(\nu h_{11/2})^{-1};8^-\rangle$ (31.7\%)
and $|(\pi g_{7/2})^3(\pi d_{5/2})^2(\nu h_{11/2})^{-1};8^-\rangle$ (35.2\%).
These dominant components in the $5^+$ and $8^-$ states
cannot be connected with an E3 transition operator
and hence do not contribute to the isomer decay.
A small fraction of the ground state, however, involves
either a neutron-hole in $\nu d_{5/2}$ 
,{\it i.e.}, $|(\pi g_{7/2})^5(\nu d_{5/2})^{-1};5^+\rangle$ (0.57\%)
and $|(\pi g_{7/2})^3(\pi d_{5/2})^2(\nu d_{5/2})^{-1};5^+\rangle$ (0.25\%),
or a neutron-hole in $\nu g_{7/2}$ 
,{\it i.e.}, $|(\pi g_{7/2})^5(\nu g_{7/2})^{-1};5^+\rangle$ (0.45\%)
and $|(\pi g_{7/2})^3(\pi d_{5/2})^2(\nu g_{7/2})^{-1};5^+\rangle$ (0.25\%).
Because these very small components are responsible for the E3 transition,
the corresponding $B$(E3) value is strongly suppressed compared to the 
single-particle estimate discussed above.
The full shell-model calculation yields
\begin{equation}
B({\rm E3};8_1^-\rightarrow5_{\rm gs}^+)=
0.40~e^2{\rm fm}^6=
3.6\cdot10^{-4}~{\rm W.u.},
\label{e_be3}
\end{equation}
if the ``standard'' values for the effective charges are taken,
$e_\nu=0.5e$ and $e_\pi=1.5e$.
In the same way one obtains for the M4 transition
of the isomer to the first-excited $4^+$ level
\begin{equation}
B({\rm M4};8_1^-\rightarrow4_1^+)=
0.408\cdot10^6~\mu_{\rm N}^2{\rm fm}^6=
12.6~{\rm W.u.},
\label{e_bm4}
\end{equation}
where unquenched values of $g$-factors have been used.

From the analysis of the single-particle estimates presented above
a strong dependence of the E3 transition rate
on the neutron effective charge $e_\nu$ can be expected.
This is confirmed by taking the effective charges
$e_\nu=0$ and $e_\pi=1.5e$,
in which case the following result is obtained:
\begin{equation}
B({\rm E3};8_1^-\rightarrow5_{\rm gs}^+)=
0.0068~e^2{\rm fm}^6=
6.1\cdot10^{-6}~{\rm W.u.},
\label{e_be3b}
\end{equation}
which leads to a partial half-life of $T_{1/2}({\rm E3})=17.5$~s
(with the conversion coefficient $\alpha_{\rm E3}=0.02$).
This results in a total (E3+M4) half-life of 10~s,
reasonably close to the observed value.

We have estimated the branching ratio of the 518-keV level
without correcting for internal conversion
and using the calculated gamma transition probabilities,
adopting a zero neutron effective charge, $e_{\nu}=0$.
The value comes out to be 99.34\%
compared to 99.88\% found from experiment.
The E3/M4 mixing ratio for the 518-keV gamma,
which is defined as the ratio of the two absolute transition amplitudes,
comes out as $\sim0.2$,
again with vanishing neutron effective charge, $e_{\nu}=0$.
With this value of the mixing ratio,
one can estimate theoretically
the internal conversion coefficients for this transition from BRICC~\cite{bricc}
and they are found to be close to those obtained experimentally.

Overall, the shell model is in reasonable agreement with observation
but there are discrepancies.
Most notably, the shell model predicts
the 105-keV transition to be predominantly M1
with a very small E2 admixture.
This is at variance with the experimental observation.
Furthermore, it still remains to be understood why the neutron effective charge
would be zero for the E3 transition in this nucleus.

\section{Conclusion and Outlook}\label{sec:conclusion}
In summary, we have for the first time determined the excitation energy of 517.9(1)~keV of the $8^-$ isomeric state in $^{136}$Cs. The isomer decays with a half-life of 17.5~s via a very slow E3 transition to the ground state. 
A very small decay branch feeds a newly discovered $4^+$ state at 105 keV by an M4 transition. 
Even with this new data only three states are known in $^{136}$Cs. Coulomb excitation experiments with $^{136m}$Cs and $^{136}$Cs beams respectively, transfer reactions like $^{135}$Cs(d,p) and $^{137}$Ba(d,$^3$He) as well as neutron capture reactions $^{135}$Cs(n,$\gamma$) should be used to find new levels and shed further light on the nuclear structure
of this amazingly unexplored nucleus that is so close to stability.

\begin{acknowledgments}
This work was supported by BMBF under contracts 06ML234, 06MT238 and 06DA9036I, by the European Commission within the Sixth Framework Programme through I3-EURONS (contract no. RII3-CT-2004-506065), by HIC for FAIR and by the PhD Sandwich programme of the Embassy of France in India.
\end{acknowledgments}
\bibliography{136cs}

\begin{thebibliography}{10}
\expandafter\ifx\csname natexlab\endcsname\relax\def\natexlab#1{#1}\fi
\expandafter\ifx\csname bibnamefont\endcsname\relax
  \def\bibnamefont#1{#1}\fi
\expandafter\ifx\csname bibfnamefont\endcsname\relax
  \def\bibfnamefont#1{#1}\fi
\expandafter\ifx\csname citenamefont\endcsname\relax
  \def\citenamefont#1{#1}\fi
\expandafter\ifx\csname url\endcsname\relax
  \def\url#1{\texttt{#1}}\fi
\expandafter\ifx\csname urlprefix\endcsname\relax\def\urlprefix{URL }\fi
\providecommand{\bibinfo}[2]{#2}
\providecommand{\eprint}[2][]{\url{#2}}

\bibitem[{\citenamefont{Ravn et~al.}(1975)\citenamefont{Ravn, Sundell, and
  Westgaard}}]{ravn}
\bibinfo{author}{\bibfnamefont{H.~L.} \bibnamefont{Ravn}},
  \bibinfo{author}{\bibfnamefont{S.}~\bibnamefont{Sundell}}, \bibnamefont{and}
  \bibinfo{author}{\bibfnamefont{L.}~\bibnamefont{Westgaard}},
  \bibinfo{journal}{J. Inorg. Nucl. Chem.} \textbf{\bibinfo{volume}{37}},
  \bibinfo{pages}{383} (\bibinfo{year}{1975}).

\bibitem[{\citenamefont{Thibault et~al.}(1981)}]{thibault}
\bibinfo{author}{\bibfnamefont{C.}~\bibnamefont{Thibault}}
  \bibnamefont{et~al.}, \bibinfo{journal}{Nucl. Phys. A}
  \textbf{\bibinfo{volume}{367}}, \bibinfo{pages}{1} (\bibinfo{year}{1981}).

\bibitem[{\citenamefont{Akimov et~al.}(2005)\citenamefont{Akimov, Bower,
  Breidenbach, Conley et~al.}}]{akimov05}
\bibinfo{author}{\bibfnamefont{D.}~\bibnamefont{Akimov}},
  \bibinfo{author}{\bibfnamefont{G.}~\bibnamefont{Bower}},
  \bibinfo{author}{\bibfnamefont{M.}~\bibnamefont{Breidenbach}},
  \bibinfo{author}{\bibfnamefont{R.}~\bibnamefont{Conley}},
  \bibnamefont{et~al.}, \bibinfo{journal}{Nuc. Phys. B - Conf. Suppl.}
  \textbf{\bibinfo{volume}{138}}, \bibinfo{pages}{224} (\bibinfo{year}{2005}).

\bibitem[{\citenamefont{Suhonen and Civitarese}(1998)}]{suhonen98}
\bibinfo{author}{\bibfnamefont{J.}~\bibnamefont{Suhonen}} \bibnamefont{and}
  \bibinfo{author}{\bibfnamefont{O.}~\bibnamefont{Civitarese}},
  \bibinfo{journal}{Phys. Rep.} \textbf{\bibinfo{volume}{300}},
  \bibinfo{pages}{123} (\bibinfo{year}{1998}).

\bibitem[{bri()}]{bricc}
\emph{\bibinfo{title}{Evaluated nuclear structure data file}},
  \urlprefix\url{http://www.nndc.bnl.gov/bricc/}.

\bibitem[{\citenamefont{Baba et~al.}(1986)\citenamefont{Baba, Sekine, Hata
  et~al.}}]{baba}
\bibinfo{author}{\bibfnamefont{S.}~\bibnamefont{Baba}},
  \bibinfo{author}{\bibfnamefont{T.}~\bibnamefont{Sekine}},
  \bibinfo{author}{\bibfnamefont{K.}~\bibnamefont{Hata}}, \bibnamefont{et~al.},
  \bibinfo{journal}{JAERI Tandem Linac VDG, Ann. Rept} p. \bibinfo{pages}{144}
  (\bibinfo{year}{1986}).

\bibitem[{\citenamefont{Talmi}(1993)}]{Talmi93}
\bibinfo{author}{\bibfnamefont{I.}~\bibnamefont{Talmi}},
  \emph{\bibinfo{title}{Simple Models of Complex Nuclei. The Shell Model and
  Interacting Boson Model}} (\bibinfo{publisher}{Harwood, Chur},
  \bibinfo{year}{1993}).

\bibitem[{\citenamefont{Brussaard and Glaudemans}(1977)}]{Brussaard77}
\bibinfo{author}{\bibfnamefont{P.~J.} \bibnamefont{Brussaard}}
  \bibnamefont{and} \bibinfo{author}{\bibfnamefont{P.~W.~M.}
  \bibnamefont{Glaudemans}}, \emph{\bibinfo{title}{Shell-model Applications in
  Nuclear Spectroscopy}} (\bibinfo{publisher}{North-Holland, Amsterdam},
  \bibinfo{year}{1977}).

\bibitem[{\citenamefont{Brown et~al.}(2005)\citenamefont{Brown, Stone, Stone,
  Towner, and Hjorth-Jensen}}]{Brown05}
\bibinfo{author}{\bibfnamefont{B.~A.} \bibnamefont{Brown}},
  \bibinfo{author}{\bibfnamefont{N.~J.} \bibnamefont{Stone}},
  \bibinfo{author}{\bibfnamefont{J.~R.} \bibnamefont{Stone}},
  \bibinfo{author}{\bibfnamefont{I.~S.} \bibnamefont{Towner}},
  \bibnamefont{and}
  \bibinfo{author}{\bibfnamefont{M.}~\bibnamefont{Hjorth-Jensen}},
  \bibinfo{journal}{Phys. Rev. C} \textbf{\bibinfo{volume}{71}},
  \bibinfo{pages}{044317} (\bibinfo{year}{2005}).

\bibitem[{\citenamefont{Brown et~al.}(2007)}]{Brownun}
\bibinfo{author}{\bibfnamefont{B.~A.} \bibnamefont{Brown}}
  \bibnamefont{et~al.}, \bibinfo{journal}{MSU-NSCL report}
  (\bibinfo{year}{2007}).

\end{thebibliography}

\end{document}